\documentclass[12pt]{article}

\usepackage[T1]{fontenc}

\usepackage{amsmath}
\usepackage{amssymb}
\usepackage{amsfonts}
\usepackage{graphicx}
\usepackage[dvipsnames]{xcolor}
\usepackage[colorlinks=true,urlcolor=blue,linkcolor=blue,citecolor=blue]{hyperref}
\usepackage{cite}

\usepackage[top=1.135in,bottom=1.17in,left=1.16in,right=1.16in]{geometry}
\numberwithin{equation}{section}

\allowdisplaybreaks

\newcommand{\ii}{{\rm i}}
\renewcommand{\d}{{\partial}}

\newcommand{\eq}[1]{\begin{align}#1\end{align}}
\newcommand{\nn}{\nonumber}

\newcommand{\lag}{{\cal L}}
\newcommand{\hc}{{\rm h.c.}}

\begin{document}
\thispagestyle{empty}

\begin{center}
{\bf\Large \boldmath
The $\rho$ parameter and $H^0\to \ell_i \ell_j$ in models with TeV sterile neutrinos}

\vspace{50pt}

Gerardo~Hern\'andez-Tom\'e, Jos\'e~I.~Illana, Manuel~Masip 

\vspace{16pt}

{\it CAFPE and Departamento de F{\'\i}sica Te\'orica y del Cosmos} \\
{\it Universidad de Granada, E-18071 Granada, Spain}

\vspace{16pt}

{\tt ghernandezt@correo.ugr.es, jillana@ugr.es, masip@ugr.es}

\today

\vspace{30pt}

\end{center}

\begin{abstract}
The presence of massive sterile neutrinos $N$ mixed with the active ones induces flavor violating processes in the charged lepton sector at the loop level. In particular, the amplitude of $H^0\to \bar{\ell}_i\ell_j$ is expected to be proportional to the product of heavy-light Yukawa couplings $y_iy_j=2 \,s_{\nu_i} s_{\nu_j}\, m_N^2/v^2$, where $s_{\nu_{i,j}}$ express the heavy-light neutrino mixings. Here we revisit these Higgs decays in the most generic extension of the neutrino sector, focusing on large values of $y_{i}$. We show that decoupling effects and a cancellation between the two dominant contributions to these processes makes the amplitude about one hundred times smaller than anticipated. We find that perturbative values of $y_{i}$ giving an acceptable contribution to the $\rho$ parameter imply ${\cal B}(H^0\to \bar{\ell}_i\ell_j)<10^{-8}$ for any lepton flavors, a rate that is not accessible at current colliders.

\end{abstract}

\newpage

\section{Introduction}
The nature of the neutrino masses remains as one of the most intriguing questions in particle physics. 
Neutrinos are {\it different} from the other fermions in that the $SU(2)_L$ singlet required to give them an electroweak (EW) mass is not protected by chirality. The possible mass of this singlet will then define a new scale that, if very large, would explain the tiny value of the neutrino masses ($m_\nu<1$ eV) deduced from flavor oscillations. Indeed, the so called type-I seesaw mechanism provides a minimal and very appealing way to complete the lepton sector of the SM. 

There are, however, other non-minimal possibilities that may be considered as well. 
Notice that gauge singlets, if present, can have {\it any} mass. From a phenomenological point 
of view, the origin of their interactions are arbitrary 
Yukawa couplings that mix them with the active neutrinos, so they could be very weakly coupled to 
matter and thus easily avoid all experimental bounds. From a model building point of view, 
they appear naturally in extensions of the SM with a cutoff much lower than the seesaw scale. 
This is the case, for example, in little 
Higgs models \cite{delAguila:2005yi,delAguila:2017ugt,delAguila:2019mvp}, 
TeV gravity models \cite{ArkaniHamed:1998rs,Randall:1999ee} or composite Higgs models \cite{Coito:2019wte}, where neutrino masses must be explained 
relying on physics at or below the TeV scale. In the end, it is the data on neutrino oscillations and charged-lepton flavor physics what decides about the motivation for these sterile neutrinos.

The appearance of non-EW terms in the extended neutrino mass matrix 
and the different gauge charges of active and sterile neutrinos will imply that the rotation 
defining the mass eigenstates does not diagonalize, respectively,
the Higgs nor the $Z$ coupling to the neutrinos. At the loop level these
flavor-changing neutral currents (FCNC), and also the charged currents coupled to the $W$ boson, induce flavor
violating processes involving the charged leptons (cLFV) \cite{Pilaftsis:1991ug,Ilakovac:1994kj,DiazCruz:1999xe,Illana:2000ic,Dinh:2012bp,Dinh:2013vya,Lindner:2016bgg,Coy:2018bxr,BhupalDev:2012zg}.
Here we will be interested in these processes. 
In particular, we will study
the cLFV decays $H^0\to\bar{\ell}_i\ell_j$  in the presence of the generic heavy sterile neutrinos that
appear in the context of low-scale seesaw models. These decay channels are currently searched 
at the LHC; at $95\%$ C.L., ATLAS \cite{Aad:2019ugc}
and CMS \cite{Khachatryan:2016rke,Sirunyan:2017xzt}  find 
\eq{
{\cal B}(H^0\to \mu e)&<  6.1 \times 10^{-5}\; {\rm (ATLAS)};\;3.5\times 10^{-4}\;{\rm (CMS)},\nonumber \\
{\cal B}(H^0\to \tau e)&< 2.8 \times 10^{-3}\; {\rm (ATLAS)};\;6.1\times 10^{-3}\;{\rm (CMS)},\nonumber\\
{\cal B}(H^0\to \tau\mu)&< 4.7 \times 10^{-3}\; {\rm (ATLAS)};\;2.5\times 10^{-3}\;{\rm (CMS)},
}
where $H^0\to \ell_i\ell_j$ stands for $H^0\to \bar{\ell}_i\ell_j,\,\ell_i\bar{\ell}_j$.
Our objective is to establish the maximum rate for these processes that could possibly
be caused by the heavy sterile neutrinos. Previous literature reports approximate
results \cite{Pilaftsis:1992st,Herrero-Garcia:2016uab,Marcano:2019rmk}
or detailed computations \cite{Arganda:2004bz,Arganda:2014dta,Thao:2017qtn,Arganda:2017vdb}
in the context of inverse seesaw models for neutrino masses. Here we will introduce a minimal
set-up \cite{Hernandez-Tome:2019lkb} that contains just two heavy neutrinos but that is able to capture all the flavor effects relevant in these
processes. The simplicity of the parametrization lets us understand the limit with large (top-quark like)
Yukawa couplings for the singlets, where one may expect branching ratios near the 
current bounds. We show that the contribution from such couplings to the $\rho$ parameter 
may be acceptable (actually, we find 
remarkable that $\Delta \rho$ from the singlet fermions may have any sign), but that 
the appearance of a cancellation and of decoupling effects 
push the decay modes well below these bounds. 

 \section{The set-up}

Flavor oscillation experiments are able to access the tiny value of the neutrino masses by combining
two very different scales, $L^{-1} E_\nu \approx \Delta m_\nu^2$. In cLFV experiments, however,
the lowest available scale is $m_\ell$, so these experiments are not sensitive to $m_\nu$. Any
observable effects will then depend on the possibly much larger masses of additional fermion singlets that 
mix with the active flavors. It turns out that to capture all the cLFV effects in a consistent way it 
will suffice to consider two massive 2-spinors that may be defining a single Dirac fermion or two 
Majorana fields of different mass. Although these singlets will not be 
responsible for the masses of the active neutrinos, 
the key point is that all the extra ingredients required to complete the neutrino sector will have no effect
on cLFV observables.

Let us be more specific (see \cite{Hernandez-Tome:2019lkb} for details).

Consider five Majorana (self-conjugate) fields $\chi_i=\chi_{Li}+(\chi_{Li})^c$  whose left-handed component $\chi_{Li}$ includes the three active neutrinos ($i=1,2,3$) plus two sterile spinors of opposite lepton number ($i=4,5$).
We will assume that in the basis of the charged-lepton mass eigenstates the only new terms in the Lagrangian are 
\eq{
-\lag\supset&\sum_{i=1}^3\, y_i \, \tilde\Phi^\dagger \,\overline {\chi}_5 P_ L L_i + 
M\,\overline {\chi}_5 
P_ L \chi_4 + {1\over 2} \mu \,\overline {\chi}_5 P_ L
 \chi_5 + {\rm h.c.}
\label{lag}
}
Once the SM Higgs  doublet $\Phi$ gets a v.e.v.  ($\tilde\Phi=i\sigma_2\Phi^*$) the Majorana mass matrix for the 5 flavors reads
\eq{
{\cal M} = \begin{pmatrix}
        0   & 0   & 0   & 0 & m_1 \\
        0   & 0   & 0   & 0 & m_2 \\
        0   & 0   & 0   & 0 & m_3 \\
        0   & 0   & 0   & 0 & M   \\
        m_1 & m_2 & m_3 & M & \mu 
\end{pmatrix}.
\label{massmatrix}
}
Notice that we have ordered the fields according to the lepton number (L)
of their left-handed component ---positive for the first four neutrinos---, 
that $m_i$ and $M$ are Dirac masses ---entries $m'_i$ in the fourth row/column
would break L--- and that $\mu$, a Majorana mass term for the neutrino with L$(\chi_{L5})=-1$, is the only source of L-breaking in this matrix.\footnote{Notice that in inverse seesaw models the usual ordering of the two massive neutrinos is the opposite, ({\it i.e.}, first the
neutrino with ${\rm L}(\chi_L)=-1$. This ordering would imply the exchange of the 4th and 5th columns/rows in our matrix ${\cal M}$.} 
Its diagonalization  yields two states $N_{1,2}$ of mass 
\eq{
m_{N_1}&=
{1\over 2} \left(\sqrt{4\left( m_1^2+m_2^2+m_3^2+M^2\right) + \mu^2} - \mu \right) , \nonumber \\
m_{N_2}&=
{1\over 2} \left(\sqrt{4\left( m_1^2+m_2^2+m_3^2+M^2\right) + \mu^2} + \mu \right) ,
}
plus three massless neutrinos $\nu_i$. It is straightforward to find that these three neutrinos
have a component along the (2-dim) sterile flavor space (a heavy-light mixing) 
\eq{
s_{\nu_i}= {m_i\over \sqrt{m_{N_1} m_{N_2}}}.
}
For $\mu=0$ the two massive modes will define a Dirac field ($m_{N_1}=m_{N_2}$); in this case, a small 
entry $\mu'$ in position ${\cal M}_{44}$ would give a mass $m_\nu\approx \mu' (m/M)^2$ to one
of the standard neutrinos, as proposed in inverse-seesaw models \cite{Mohapatra:1986bd,Bernabeu:1987gr}. In the opposite limit, 
if $M=0$ and $\mu\to 10^{10}$ GeV the configuration describes a type-I seesaw mechanism, with 
one of the active neutrinos massive, $m_{N_1}\approx (m_1^2+m_2^2+m_3^2)/\mu$, 
while the second singlet ($\chi_4$) is massless but decoupled.
For $\mu$ in the TeV range, as long as 
$M>10\sqrt{m_1^2+m_2^2+m_3^2}$ ({\it i.e.}, the mixings are below 0.1) the model may be viable. 
At any rate, 
${\cal M}$ is a rank-2 matrix with three zero mass eigenvalues.
As we argued above, the extra spinors and couplings required 
to generate light neutrino masses will have no effect on cLFV observables. 
In particular, the so called TeV type-I seesaw models \cite{Pilaftsis:1991ug}
can be obtained by adding a third singlet with an
$O({\rm TeV})$ Majorana mass $\Lambda$; in a certain basis all these models are reduced
to the texture
\eq{
{\cal M'} = \begin{pmatrix}
        0   & 0   & 0   & \cdot & m_1 & \cdot \\
        0   & 0   & 0   & \cdot & m_2 & \cdot \\
        0   & 0   & 0   & \cdot & m_3 & \cdot \\
         \cdot  & \cdot   &  \cdot  & \cdot  & M   & \cdot \\
        m_1 & m_2 & m_3 & M & \mu & 0 \\
         \cdot &  \cdot & \cdot  & \cdot & 0   & \Lambda 
\end{pmatrix},
\label{general}
}
where the dots indicate very small entries that are necessary to generate standard
neutrino masses and  light-light
mixings but have no effect on the heavy-light mixings: any $O({\rm GeV})$  
term there would increase the rank 3 of this $6\times 6$ matrix and 
imply a non-acceptable mass spectrum.
Notice also that the third 
singlet does not introduce significant heavy-light mixings. 
Therefore, the 5 mass parameters in ${\cal M}$ (or two heavy masses plus three heavy-light mixings) 
are enough to describe all cLFV effects caused by heavy Dirac or Majorana singlets mixed 
with the three active families.

One should also stress, however, that if $\mu\not= 0$ the matrix above is not stable under radiative corrections \cite{Bolton:2019pcu}: the breaking of lepton number will contribute to all the entries in ${\cal M}$ at the
loop level, which would give mass to a linear combination of the three $\nu_i$. 
If this breaking is {\it small} the mass will be acceptable ({\it i.e.}, below 1 eV), but if $\mu$ is large
the model will require a fine tuned cancellation of these loop contributions. In summary, the 
texture that we propose in Eq.~(\ref{massmatrix}) must be understood as approximate and 
established at the loop level  where we work. Despite the fine tune that this involves,
we will consider TeV values of $\mu$ in order 
to understand the genuine Majorana effects on cLFV observables and on the contribution to 
the $\rho$ parameter from heavy singlets.

\section{Large Yukawa couplings and $\Delta \rho$}

The Yukawa couplings $y_i$ in Eq.~(\ref{lag}) are the origin of any interactions of the
heavy singlets, and the rate of  $H^0\to\bar{\ell}_i\ell_j$ will certainly grow with them. In our model,
their relation with the masses and mixings is
\eq{
y_{i} =\sqrt{2} \;\frac{m_i}{v}=\sqrt{2} \;\frac{\sqrt{m_{N_1}m_{N_2}}}{v}\, s_{\nu_i}\,.
}
The expression above shows that, for a fixed value of the mixings consistent with current constraints, large singlet masses will probe large values of $y_i$.  These couplings,
however, break the custodial symmetry of the SM and will contribute to the $\rho$ parameter 
(or to the Peskin-Takeuchi parameter
$T=(\rho-1)/\alpha$). These oblique corrections can be easily obtained from the 
contribution of the heavy neutrinos to the gauge boson self-energies at $q^2=0$,
\eq{
\Delta \rho = \frac{ \left(\Pi_{WW}\right)_{N_{1,2}}}{M_W^2}-\frac{ \left(\Pi_{ZZ}\right)_{N_{1,2}}}{M_Z^2},
}
and they are constrained to be $|\Delta \rho | \lesssim 0.0005$ \cite{Tanabashi:2018oca}. 
At one loop and neglecting charged lepton masses, we find 
(see the couplings to gauge and Goldstone bosons in Appendix~\ref{FR})
\eq{
\Delta\rho &= \frac{g^2}{32\pi^2 M_W^2} 
\Big(\sum_{k=1}^3 s_{\nu_k}^2\Big)^2 
\frac{m_{N_1}^2m_{N_2}^2}{(m_{N_1}+m_{N_2})^2}
\left(3-2\frac{m_{N_1}^2+m_{N_2}^2-m_{N_1}m_{N_2}}{m_{N_2}^2-m_{N_1}^2}
\ln\frac{m_{N_2}}{m_{N_1}}\right).
\label{rho}
}
This result presents some interesting features. Let us assume for simplicity mixing
with just $\nu_\tau$ and consider first the
case with a Dirac singlet field ($\mu=0$). The contribution  is then obtained from Eq.~(\ref{rho})
by taking the limit $m_{N_1},m_{N_2}\to m_{N}$:
\eq{
\Delta\rho = \frac{g^2}{64\pi^2 M_W^2} 
\, s_{\nu_\tau}^4 m_{N}^2 = \frac{g^2}{64\pi^2 M_W^2} 
\, s_{\nu_\tau}^2 \left(y_3 {v\over \sqrt{2}} \right)^2.
\label{rhoDirac}
}
If we compare this with the correction from the top quark,
\eq{
\Delta\rho_{t} = 3\,\frac{g^2}{64\pi^2 M_W^2}\left(y_t {v\over \sqrt{2}} \right)^2
\simeq 0.009,
}
we see an extra suppression by a decoupling factor
of $s_\nu^2$. Obviously, if the heavy neutrino were a sequential doublet with a purely 
EW mass this suppression would be absent; in this case the contribution should be canceled by 
restoring the custodial symmetry with 
a very similar Yukawa coupling of the charged lepton in the same doublet. But here, for
$s_{\nu_\tau}<0.1$ and $y_3<\sqrt{4\pi}$ we have that $\Delta\rho<0.00038$ is within
the experimental bounds.

Another interesting limit goes in the opposite direction: a Majorana mass $\mu$ much larger
than $M$ and then $m_{N_2}\gg m_{N_1}$. It is easy to see that if 
\eq{
m_{N_2}^2 > 30 \,m_{N_1}^2 \qquad (\mbox{or } \mu > 2.1 \,M)
}
the second term in Eq.~(\ref{rho}) dominates and the contribution to $\Delta\rho$ is 
negative, something remarkable as multiplets of non-degenerate Dirac  fermions 
always give $\Delta\rho>0$. For $s_{\nu_\tau}<0.1$ and $y_3<\sqrt{4\pi}$ we obtain 
$-\Delta\rho<0.00012$. The correction for a type-I seesaw mechanism ($M=0$, 
$\mu\gg 1$ TeV) is just
\eq{
\Delta\rho \approx  -\frac{g^2}{32\pi^2 M_W^2} 
\,m_{\nu_\tau}^2 \left( 2\,\ln\frac{\mu}{m_{\nu_\tau}}-3\right),
\label{rhoMajorana}
}
with $m_{\nu_\tau}=y_3^2\, v^2 /(2\mu)$. Our results for $\Delta \rho$
from TeV fermion singlets are consistent with the generic ones in \cite{Einhorn:1981cy}.

\section{\boldmath$H^0\to \bar{\ell_i}\ell_j$}

\begin{figure*}
\begin{center}
\begin{tabular}{cccc}
\includegraphics[scale=0.65]{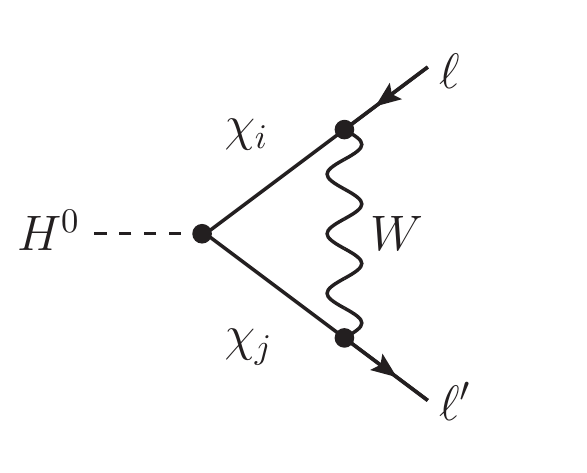} &
\includegraphics[scale=0.65]{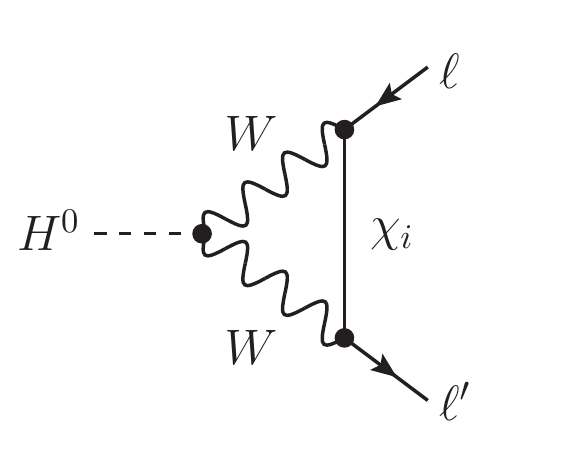} &
\includegraphics[scale=0.65]{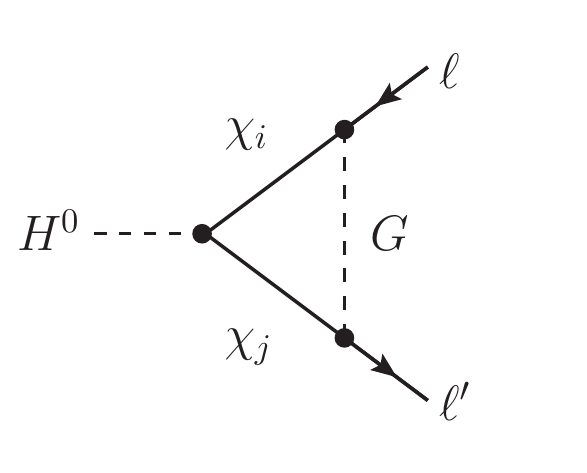} &
\includegraphics[scale=0.65]{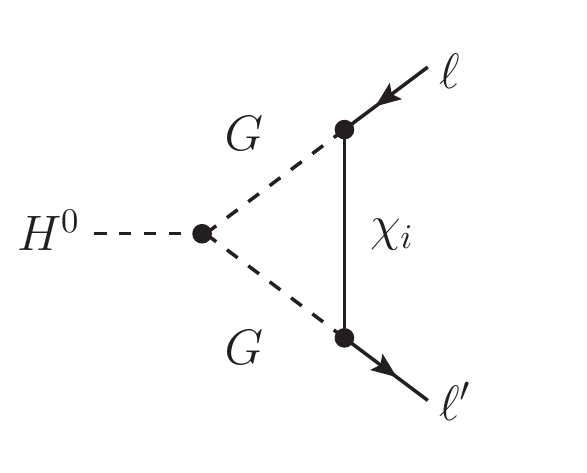} \\
$W\chi\chi$ & $\chi WW$ & $G\chi\chi$ & $\chi GG$ \\[1ex]
\includegraphics[scale=0.65]{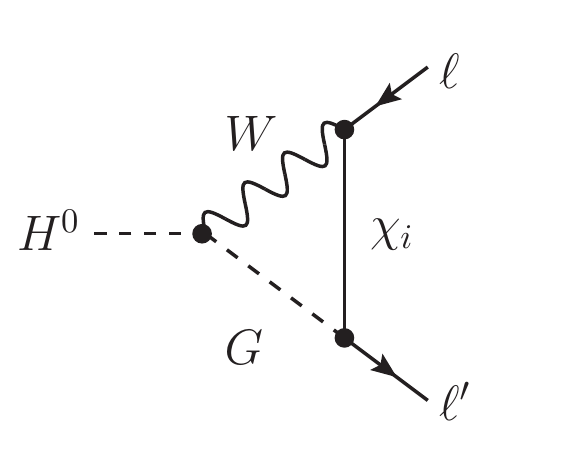} &
\includegraphics[scale=0.65]{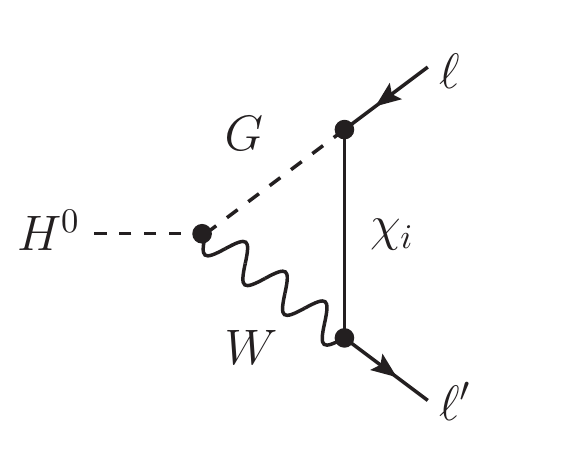} &
\includegraphics[scale=0.65]{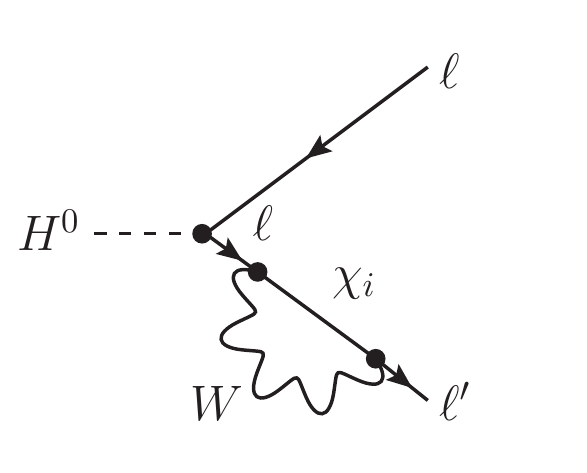} &
\includegraphics[scale=0.65]{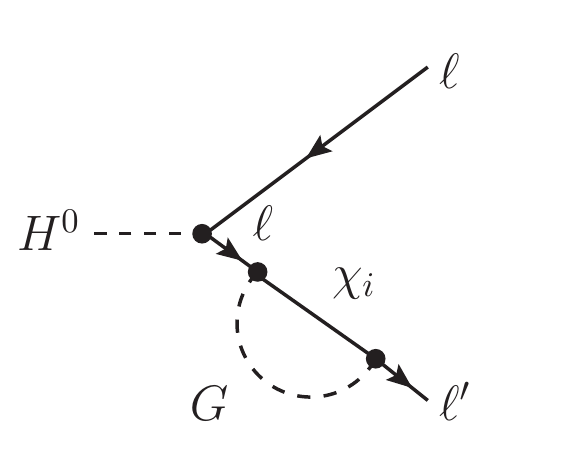} \\
\multicolumn{2}{c}{$\chi WG$} & $W\chi$ & $G\chi$  
\end{tabular}
\end{center}
\caption{Diagrams contributing to $H^0\to\bar{\ell}\ell'$ in the Feynman-'t Hooft gauge, for $m_{\ell'}=0$. Diagram $W\chi$ is proportional to $m_\ell^3$ and will be neglected.
\label{fig:diagrams}}
\end{figure*}

The one-loop amplitude for $H^0\to \bar{\ell}_i\ell_j$ 
is mediated in the Feynman-'t Hooft gauge by the diagrams in Fig.~\ref{fig:diagrams}.
One can see that all these diagrams are proportional to
\eq{
y_i \,y_j \,y_{\ell_i}
=2\sqrt{2}\, s_{\nu_i} s_{\nu_j} \,{m_{N_1} m_{N_2}m_{\ell_i}\over v^3}\,,
\label{yukawa}
}
where $\ell_i$ above refers to the heavier final lepton. In addition, diagrams $W\chi\chi$, $\chi WG$ and
$W\chi$ are proportional to $g^2$, $\chi WW$ is proportional to $g^4$, and $\chi GG$ to 
the Higgs quartic coupling $\lambda$. Of course, each diagram will also depend on the
mass and spin of the particles inside the loop, but one may expect that $G\chi\chi$ and
$G\chi$ dominate with a contribution of order ${\cal M}\approx y_i \,y_j \,y_{\ell_i}/(16\pi^2)$. 
This estimate coincides with what is expected using an effective field theory approach (see Ref.~\cite{Herrero-Garcia:2016uab}).

Using this estimate, we can deduce the maximum branching ratio in Higgs decays by 
comparing with ${\cal B}(H^0\to \bar{b}b)\simeq0.6$. For the decay
$H^0\to \tau e$ 
we expect
\eq{
{\cal B}(H^0\to \tau e) &= {\cal B}(H^0\to \bar{b}b)\,
\frac{2\,\Gamma(H^0\to \bar{\tau} e)}{\Gamma(H^0\to\bar{b}b)} \nonumber\\
&\approx {\cal B}(H^0\to \bar{b}b)\,
{2\over 3} \left(  {y_3\,y_1 \,y_{\tau}\over y_b\, 16\pi^2 } \right)^2.
\label{estimate}
}
Taking $y_i <\sqrt{4\pi}$ this  gives ${\cal B}(H^0\to \tau e)<4\times 10^{-4}$, a value
that could be accessible once the LHC reaches its highest luminosity. However, a  
precise calculation will show that this is not the case.

First of all, although their sum is finite, the diagrams $G\chi\chi$ and
$G\chi$ are both divergent. In addition, there is a value of the heavy
neutrino mass that exactly cancels the sum of both contributions. For $m_{N_1}=m_{N_2}$ this is 
\eq{
\widetilde m_N\approx 0.57\, \frac{M_H}{\sqrt{s_{\nu_e}^2+s_{\nu_\mu}^2+s_{\nu_\tau}^2}}.
\label{mass}
}
Finally, at masses of the heavy neutrinos above $\widetilde m_N$ there are decoupling effects, like the extra factor
of $s_{\nu_\tau}^2$ in $\Delta\rho$ found in the previous section. 

Let us be more definite. We write the decay amplitude
 \eq{
{\cal M}(H^0\to \bar {\tau} e)= \bar u(p_2) \,{f^{\tau e}\over v} \left[ m_{\tau} P_R+
m_{e} P_L \right] v(p_1)
}
and will give the results in terms of  $m_{N_1}$ and the ratio
\eq{
r\equiv {m_{N_2}^2\over m_{N_1}^2} \ge 1\,.
}
Constraints from flavor-diagonal processes  \cite{delAguila:2008pw,deBlas:2013gla,Fernandez-Martinez:2016lgt,Coutinho:2019aiy}
together with 
\eq{
{\cal B}(\mu\to e\gamma)\approx \frac{3\alpha}{8\pi} s^2_{\nu_\mu} s^2_{\nu_e} < 4.2\times 10^{-13}
}
imply \cite{Hernandez-Tome:2019lkb}
\eq{
s_{\nu_e}^{\rm max} = 0.05, \quad
s_{\nu_\mu}^{\rm max} = 4.5\times10^{-4}, \quad
s_{\nu_\tau}^{\rm max} = 0.075.
\label{maxmix}
}

\begin{figure}[t]
\begin{center}
\includegraphics[scale=0.6]{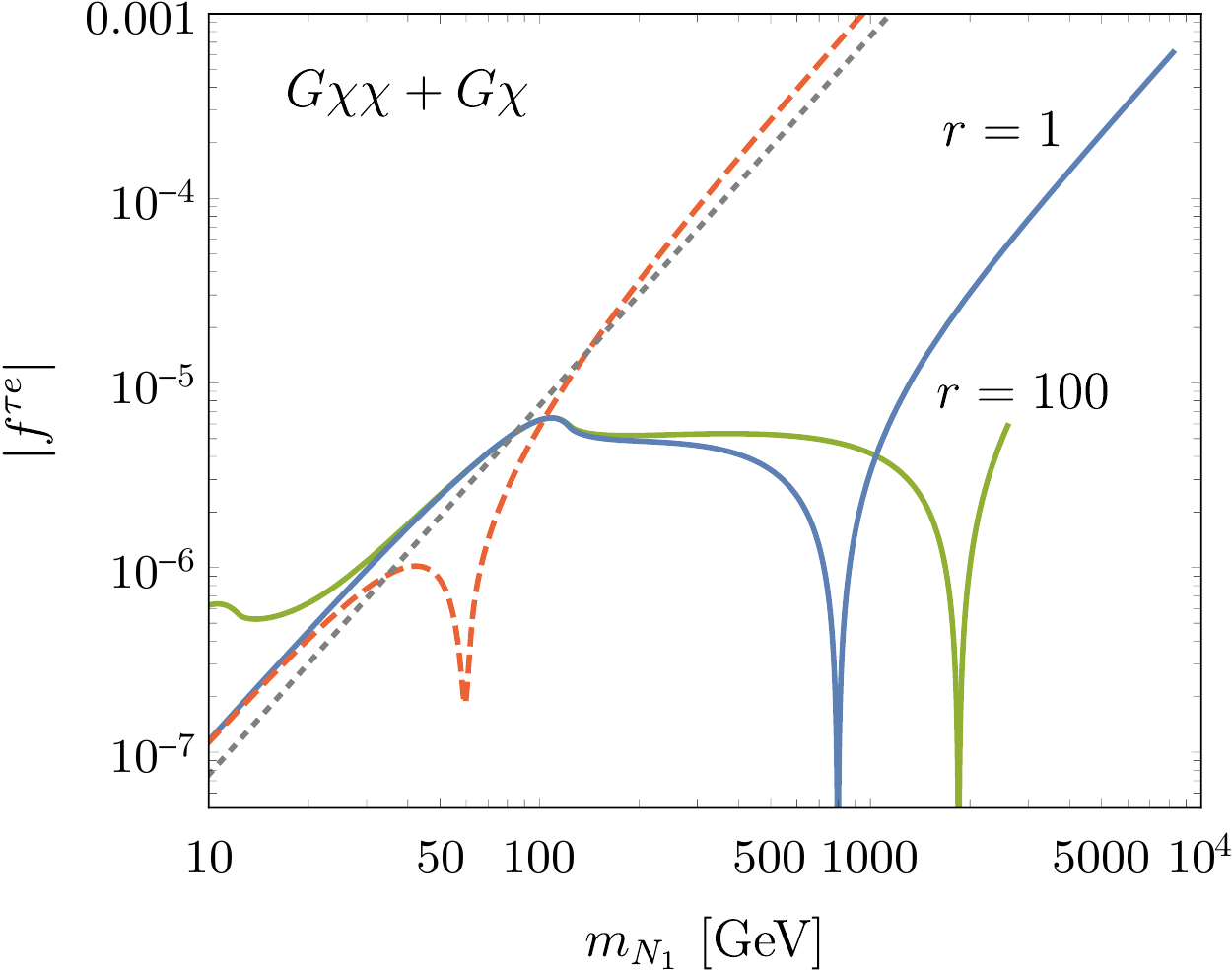}
\caption{Contribution to $|f^{\tau e}|$ from the dominant diagrams $G\chi\chi+G\chi$ 
for fixed (maximal) mixings and different heavy neutrino masses (notice that Yukawa couplings grow with the mass). UV divergences cancel in $G\chi\chi+G\chi$. The blue line ($r=1$) shows the heavy Dirac case whereas the red line ($r=100$) corresponds to two Majoranas with $m_{N_2}=10\,m_{N_1}$. We have included the estimate of $|f^{\tau e}|$ in Eq.~(\ref{estimate}) for $r=1$ (gray dots) as well as the contribution from a massive neutrino with an active left-handed component (red dashes).
\label{fig2}}
\end{center}
\end{figure}

In Fig.~\ref{fig2} we plot the contribution to $|f^{\tau e}|$ from the  $G\chi\chi+G\chi$
for these maximal mixings and $r=1,100$.
We see that it grows with the heavy-light Yukawas (with the size anticipated below Eq.~(\ref{yukawa})), then there appears the cancellation 
at $\widetilde m_N$
discussed above, and finally the amplitude reaches a regime where it grows again with the Yukawas
but is suppressed by a (decoupling)
factor of $s_{\nu_e}^2+s_{\nu_\mu}^2+s_{\nu_\tau}^2\approx 0.01$ for maximal mixings. 
This suppression is consistent with the results obtained in \cite{Arganda:2017vdb} using the mass insertion approximation in the region where Yukawa couplings become dominant ($y>g,\lambda$). The curves in Fig.~\ref{fig2} finish at
$y_i=\sqrt{4\pi}$, that imply $m_{N_1}=8.2\;(2.6)$~TeV for $r=1\;(100)$.
The plot also shows that 
Majorana effects 
($r=1$ gives a Dirac heavy neutrino) do not change the qualitative behavior of the amplitude
and are not able to increase the maximum value of $|f^{\tau e}|$.
In the same plot we have included the amplitude for a heavy neutrino in a $SU(2)_L$ 
doublet:\footnote{This case requires a charged lepton of similar mass to cancel $\Delta \rho$ as well as extra EW fermions to cancel anomalies ({\it e.g.}, to complete the whole 
{\it sequential} 4th family) that are
excluded by the LHC.}
a Dirac field with an active left handed component. Such a neutrino does not
decouple for large values of $m_N$, which is purely EW; the plot reveals that 
in this case $G\chi\chi+G\chi$ 
follows the scaling in Eq.~(\ref{estimate}) for all values of the heavy neutrino mass.
The origin of the suppression proportional to the squared mixings with the heavy singlets is the flavor-changing vertex $H\chi_i\chi_j$, that would be flavor diagonal if the neutrinos were active (see Appendix~\ref{FR}).

\begin{figure}[t]
\begin{center}
\includegraphics[scale=0.6]{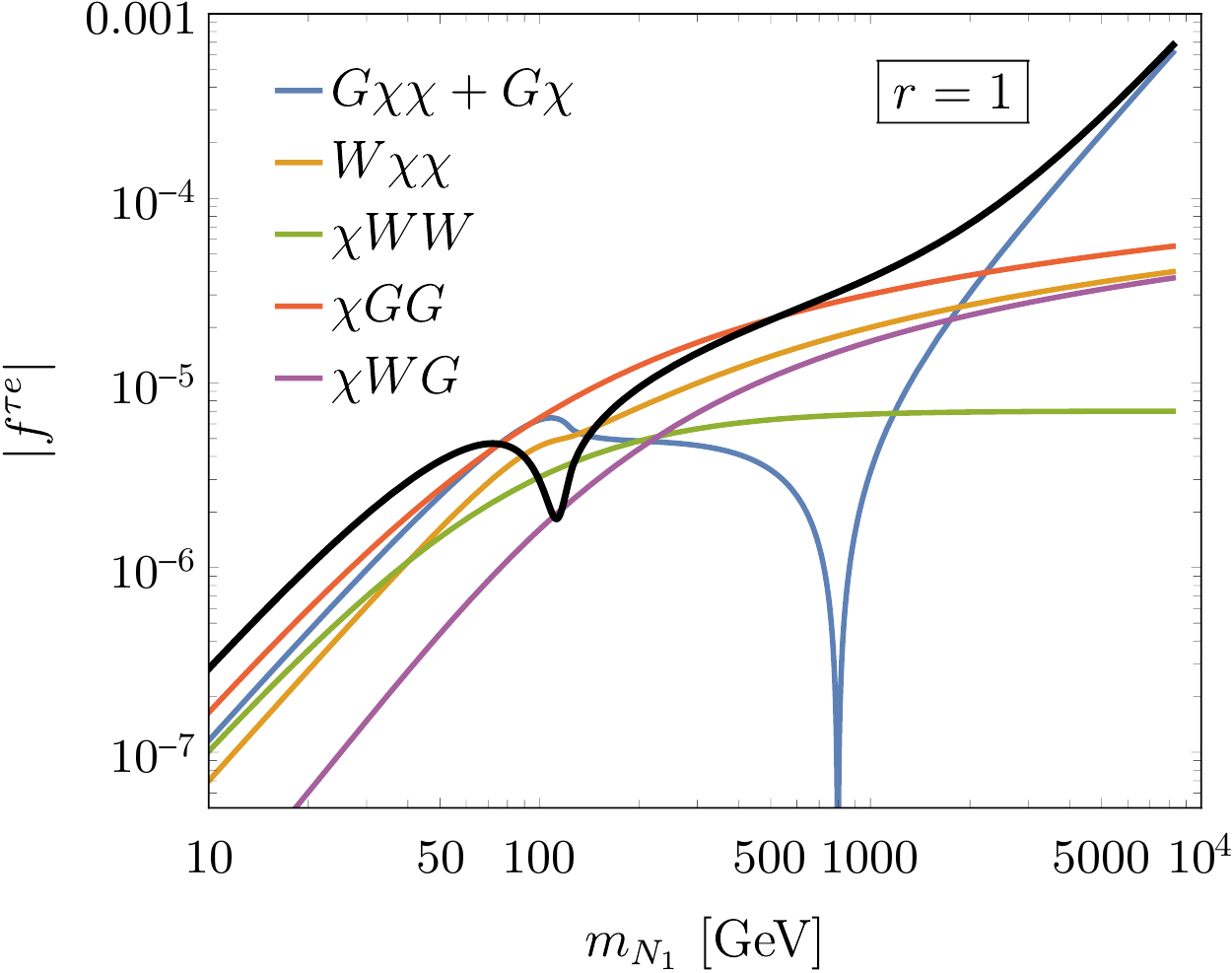}\hfill
\includegraphics[scale=0.6]{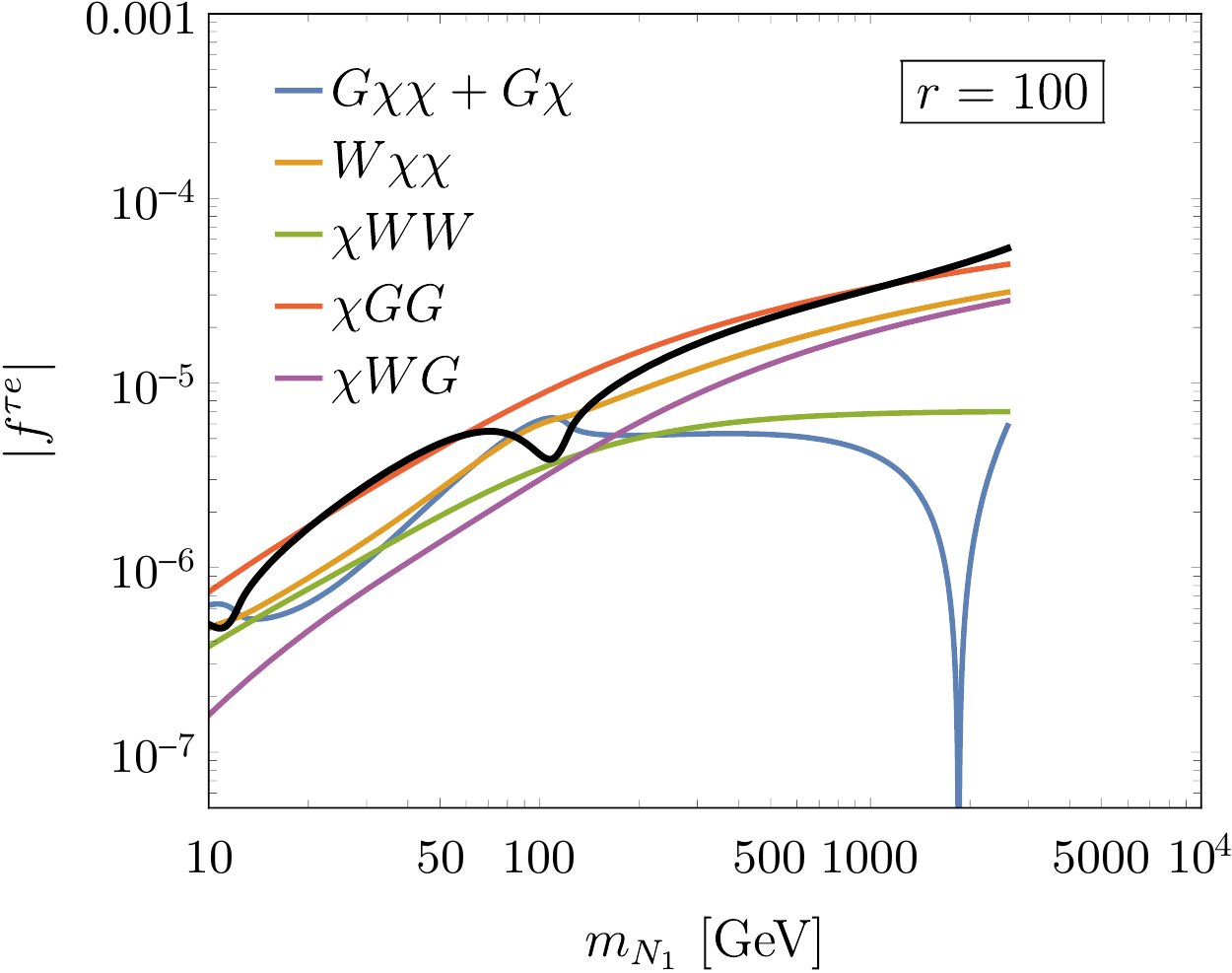}
\caption{Contribution to $|f^{\tau e}|$ from the different diagrams in Fig.~\ref{fig:diagrams} for  fixed maximal mixings and $r=1,100$. The thick line corresponds to the sum of all the diagrams. All amplitudes are real except for $G\chi\chi+G\chi$  and $W\chi\chi$, that have an imaginary part for $m_{N_1}<M_H/2$. The real  part of the amplitudes are positive except for $W\chi\chi$ in the whole mass interval and  $G\chi\chi+G\chi$, that changes sign from negative to positive at intermediate masses producing a drop.
\label{fig3}}
\end{center}
\end{figure}

In Fig.~\ref{fig3} we plot the modulus of each contribution and of the sum of all diagrams for $r=1,100$. We have considered $m_{N_1}$ from 10 GeV to its maximum perturbative value just to illustrate the behavior of each contribution, although our analysis focuses on large neutrino masses.\footnote{For $m_N\le m_Z$ current data from colliders set stringent direct limits on the active-sterile mixing from gauge boson and Higgs decays \cite{Adriani:1992pq,BhupalDev:2012zg,Das:2017rsu}.} We see that the dominant contribution comes from $\chi GG$ except at very large Yukawa couplings, {\it i.e.}, maximal mixings and heavy neutrino masses above 2 TeV, 
when diagrams $G\chi\chi+G\chi$ take the lead despite the decoupling factor, yielding a maximum value that is two orders of magnitude smaller than the naive guess given before. In Appendix~\ref{ff} we present expressions for the form factors and give further details of our computation.

\section{Summary and discussion}

Vectorlike fermions at the TeV scale are a possibility with interesting phenomenological consequences. If they are quarks or charged leptons that mix with the active families, their different EW numbers will induce tree-level FCNCs that are very constrained experimentally. If they are neutrinos, however, collider effects appear at the loop level and the bounds are weaker. Here we have focused on cLFV decays of the Higgs boson. These processes have been studied by several groups, with results that sometimes appear as contradictory. In this work we have proposed a set-up with two sterile fields that captures all flavor effects and lets us understand the results in a simple way.  The model reveals, for example, that in generic low-scale seesaw models Majorana singlets with TeV mass and unsuppressed mixings with the active neutrinos are indeed possible, although they require a fine-tuned cancellation of loop corrections so that the observed neutrinos have sub-eV masses (notice that inverse seesaw models the heavy neutrinos are quasi-Dirac). Or that large values of the heavy-light Yukawa couplings in these models have an impact on $\Delta \rho$ for large enough heavy-light mixings.

Our analysis shows that the Higgs decay modes $H^0\to \bar{\ell}_i\ell_j$ are not accessible at colliders. The rate of these decays is expected to grow with the Yukawa couplings that mix active and sterile neutrinos, but a cancellation of different contributions and decoupling effects proportional to the sum of squared mixings damp the final result. These two features are clearly shown in Fig.~\ref{fig2}. We see that for a fixed mixing and a relatively light neutrino mass the amplitude grows with the Yukawa couplings (which are proportional to the mass) as expected, until the scale in Eq.~(\ref{mass}) where the dominant amplitude goes to zero and changes sign. At heavier neutrino masses the amplitude grows again with the  couplings, however, all but a component of order $(s_{\nu_e}^2+s_{\nu_\mu}^2+s_{\nu_\tau}^2)^{1/2}$ is decoupled: the amplitude ${\cal M}\approx y_i \,y_j \,y_{\ell_i}/(16\pi^2)$ at low singlet masses becomes
of order $(s_{\nu_e}^2+s_{\nu_\mu}^2+s_{\nu_\tau}^2) y_i \,y_j \,y_{\ell_i}/(16\pi^2)$ in this decoupled regime.
As a consequence, we find that the largest branching
ratio consistent with the maximal mixings summarized in Eq.~(\ref{maxmix}) 
would correspond to the channel $H^0\to \tau e$ and is
\eq{
{\cal B}(H^0\to \tau e) < 1.4 \times 10^{-8} .
}
We conclude that the observation of cLFV in Higgs decays 
at the LHC would involve a different type of 
new physics.

\section*{Acknowledgements}

We would like to thank F. del \'Aguila, G. L\'opez-Castro, P. Roig and J. Santiago for helpful discussions.
This work was supported in part by the Spanish Ministry of Science, Innovation and Universities
(FPA2016-78220-C3, PID2019-107844GB-C21/AEI/10.13039
/501100011033), and by Junta de Andaluc{\'\i}a (FQM~101, SOMM17/6104/UGR, P18-FR-1962, P18-FR-5057).
The work of GHT has been funded by CONACYT of Mexico through the program ``Estancias postdoctorales en el extranjero 2019-2020''.

\appendix


\section{Flavor-changing vertices and mixing matrices}
\label{FR}

The neutrino mass eigenstates come from the interaction eigenstates by the replacement
\eq{
\chi_{Li} \to \sum_{j=1}^{5} U_{ij}^\nu \chi_{Lj},
\label{inter2mass}
}
where $U^\nu$ is the unitary matrix diagonalizing ${\cal M}$ (\ref{massmatrix}) into real and positive mass eigenvalues. The Lagrangian for charge-current interactions reads 
\eq{
\lag_W &= \frac{g}{\sqrt{2}} W_\mu^-  \sum_{i=1}^3 \sum_{j=1}^5
B_{ij}\bar{\ell}_i \gamma^\mu P_L \chi_j + \hc,
}
where we have used the convention $D_\mu=\d_\mu-\ii g\widetilde{W}_\mu$ for the covariant derivative and
\eq{
B_{ij}
=\sum_{k=1}^{3}\delta_{ik} U^{\nu}_{kj}
}
is a rectangular $3\times 5$ mixing matrix. In the Feynman-'t Hooft gauge one also needs
\eq{
\lag_{G^\pm} =& -\frac{g}{\sqrt{2}M_W} G^- \sum_{i=1}^3 \sum_{j=1}^5
B_{ij}\bar{\ell}_i(m_{\ell_i}P_L-m_{\chi_j}P_R) \chi_j 
+ \hc,
}
where $G^\pm$ is the charged would-be-Goldstone field. The matrix $U^\nu$ introduces tree-level flavor-changing interactions with the $Z$ and the Higgs field:
\eq{
\lag_Z =& \frac{g}{4c_W}Z_{\mu} \sum_{i,j=1}^5
\bar{\chi}_i \gamma^\mu (C_{ij} P_L - C_{ij}^* P_R) \chi_j,
\\
\lag_H =& 
-\frac{g}{4M_W} H\sum_{i,j=1}^5
\bar{\chi}_i [(m_{\chi_i}C_{ij}+m_{\chi_j}C_{ij}^*)P_L  + (m_{\chi_i}C_{ij}^*+m_{\chi_j}C_{ij})P_R] \chi_j ,
}
where
\eq{
C_{ij}=\sum_{k=1}^{3}{(U_{ki}^{\nu})}^{*}U_{kj}^{\nu}.
}
A symmetry factor of 2 must be added in the Feynman rule for vertices including two (self-conjugate) Majorana fermions \cite{Denner:1992vza,Akhmedov:2014kxa}:
\eq{
\raisebox{-0.5\height}{\includegraphics[scale=0.65]{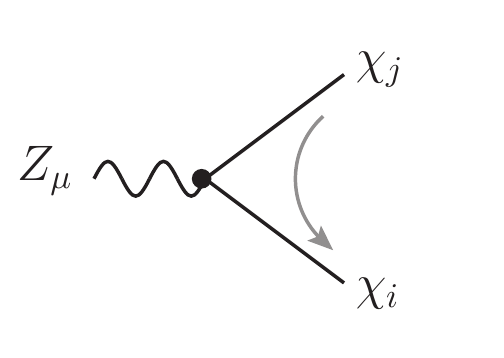}}
&\hspace{-2mm}
\ii\frac{g}{2c_W}\gamma^{\mu}\left(C_{ij}P_L -C_{ij}^{*}P_R \right),
\label{vZXX}
\\
\raisebox{-0.5\height}{\includegraphics[scale=0.65]{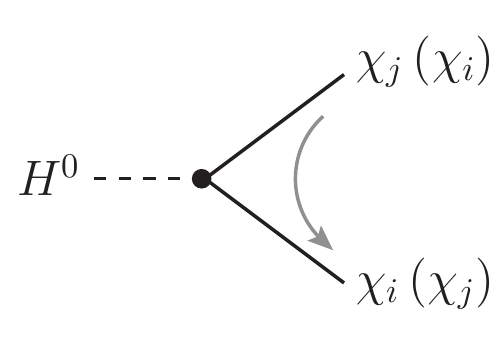}} 
&\hspace{-3mm}
-\ii\frac{g}{2M_W} 
 [(m_{\chi_i}C_{ij}+m_{\chi_j}C_{ij}^*)P_L +(m_{\chi_i}C_{ij}^*+m_{\chi_j}C_{ij})P_R].
\label{vHXX}
}
One can recover the case of active Dirac neutrinos by replacing $C_{ij}\to\delta_{ij}$, $C_{ij}^{*}\to 0$. 

The mixing matrix elements involving heavy neutrinos can be expressed in terms of heavy-light mixings and the squared mass ratio $r=m^2_{N_2}/m^2_{N_1}$ as
\eq{
B_{k N_{1}}&=-\frac{\ii\,r^{\frac{1}{4}}}{\sqrt{1+r^{\frac{1}{2}}}}s_{\nu_k}, \quad 
B_{k N_{2}}=\frac{1}{\sqrt{1+r^{\frac{1}{2}}}}s_{\nu_k},
\label{Bsfortwo}
\\
C_{N_{1}N_{1}}&=\frac{r^{\frac{1}{2}}}{1+r^{\frac{1}{2}}}\sum_{k=1}^{3}s_{\nu_k}^{2}, \quad 
C_{N_{2}N_{2}}=\frac{1}{1+r^{\frac{1}{2}}}\sum_{k=1}^{3}s_{\nu_k}^{2},
\nonumber\\
&C_{N_{1}N_{2}}=-C_{N_{2}N_{1}}=\frac{\ii\,r^{\frac{1}{4}}}{1+r^{\frac{1}{2}}}\sum_{k=1}^{3}s_{\nu_k}^2 .\label{Csfortwo}
}

\section{Form factors}\label{ff}

The form factors $f^{\ell\ell'}$ receive contributions from the one-loop diagrams of Fig.~\ref{fig:diagrams} in the Feynman-'t Hooft gauge. Neglecting charged lepton masses we find:
\eq{
f^{\ell\ell'}_{W\chi\chi} 
&=  \frac{g^2}{16\pi^2}\sum_{i,j=1}^5 B^*_{\ell i}B_{\ell' j}\,
\big\{
  C_{ij}\, \sqrt{x_i x_j}\, [c_0 + 2 c_1] \nn\\ 
& \hspace{32mm}
+ C_{ij}^*\, [x_j c_0 + (x_i+x_j) c_1]
\big\},
\label{Wchichi}\\
f^{\ell\ell'}_{\chi WW} &= \frac{g^2}{16\pi^2}\sum_{i=1}^5 B^*_{\ell i}B_{\ell' i}\,
[-2 \overline{c}_1],
\\
f^{\ell\ell'}_{G\chi\chi} 
&= \frac{g^2}{16\pi^2}\sum_{i,j=1}^5 B^*_{\ell i}B_{\ell' j}\,
\bigg\{
  C_{ij}\, \sqrt{x_i x_j} \left[\frac{1}{4} - 2c_{00}+ \frac{1}{2}(x_i+x_j)c_1 + \frac{1}{2} x_Q c_{12} \right] \nn\\ 
& \hspace{32mm}
+ C_{ij}^*\, x_j \left[\frac{1}{4} - 2c_{00}+ x_i c_1 + \frac{1}{2} x_Q c_{12} \right] 
\bigg\},
\\
f^{\ell\ell'}_{\chi GG} &= \frac{g^2}{16\pi^2}\sum_{i=1}^5 B^*_{\ell i}B_{\ell' i}\,
\left[-\frac{1}{2} x_H x_i(\overline{c}_0 + \overline{c}_1) \right],
\\
f^{\ell\ell'}_{\chi WG} &= \frac{g^2}{16\pi^2}\sum_{i=1}^5 B^*_{\ell i}B_{\ell' i}\,
\left[\frac{1}{4} - 2\overline{c}_{00} - \frac{1}{2} x_i (\overline{c}_0 + 2\overline{c}_1) + \frac{1}{2} x_Q (2\overline{c}_1 + \overline{c}_{12})\right],
\\
f^{\ell\ell'}_{W\chi} &=0,
\\
f^{\ell\ell'}_{G\chi} &= \frac{g^2}{16\pi^2}\sum_{i=1}^5 B^*_{\ell i}B_{\ell' i}\,
\frac{1}{2} x_i b_0,
\label{Gchi}
}
where we have introduced the following dimensionless functions in terms of the standard Passarino-Veltman loop functions \cite{Passarino:1978jh}:
\eq{
b_0(x_i) &\equiv B_0(0;M_W^2,x_i M_W^2) = B_0(0;x_i M_W^2,M_W^2), \\
c_{00}(x_i,x_j) &\equiv C_{00}(0,Q^2,0;M_W^2,x_i M_W^2,x_j M_W^2), \\
c_{\{0,1,12\}}&(x_i,x_j)\equiv   M_W^2 C_{\{0,1,12\}}(0,Q^2,0;M_W^2,x_i M_W^2,x_j M_W^2), \\
\overline{c}_{00}(x_i) &\equiv C_{00}(0,Q^2,0;x_i M_W^2,M_W^2,M_W^2), \\
\overline{c}_{\{0,1,12\}}&(x_i)\equiv M_W^2 C_{\{0,1,12\}}(0,Q^2,0;x_i M_W^2,M_W^2,M_W^2),
}
with $x_i\equiv m_{\chi_i}^2/M_W^2$, $x_Q\equiv Q^2/M_W^2$, $x_H\equiv M_H^2/M_W^2\approx2.4$, and $Q^2=M_H^2$ for an on-shell Higgs. We use the conventions
of \cite{Hahn:1998yk}. The functions $b_0$, $c_{00}$ and $\overline{c}_{00}$ are ultraviolet divergent but, thanks to relations between $B$ and $C$ matrix elements \cite{Hernandez-Tome:2019lkb}, the divergences in $f^{\ell\ell'}_{G\chi\chi}$ and $f^{\ell\ell'}_{G\chi}$ cancel each other, and $f^{\ell\ell'}_{\chi W G}$ is finite when summing over all neutrino states. The other diagrams are finite.

It turns out convenient to cast the contributions to the form factor (\ref{Wchichi}--\ref{Gchi}) into mixing-independent functions $F$, $G$, $H$:
\eq{
f^{\ell\ell'} &= \frac{g^2}{16\pi^2}\sum_{i,j=1}^5 B_{\ell i}^* B_{\ell' i} 
\left[
\delta_{ij} F(x_i) + C_{ij} G(x_i,x_j) + C_{ij}^* H(x_i,x_j)
\right].
}
In this way, the form factor can be expressed in terms of massive neutrinos only \cite{Hernandez-Tome:2019lkb} as:
\eq{
f^{\ell\ell'} = \frac{g^2}{16\pi^2}\sum_{i,j=1}^2 & B_{\ell N_i}^* B_{\ell' N_j}
\big\{
 \delta_{ij} [F(x_{N_i})-F(0)]
\nn\\[-2ex]
&            +\delta_{ij}[G(x_{N_i},0)+G(0,x_{N_j})-2G(0,0)]
\nn\\
&            +\delta_{ij}[H(x_{N_i},0)+H(0,x_{N_j})-2H(0,0)]
\nn\\
&+C_{N_iN_j} [G(x_{N_i},x_{N_j})-G(x_{N_i},0)-G(0,x_{N_j})+G(0,0)]
\nn\\
&+C_{N_iN_j}^* [H(x_{N_i},x_{N_j})-H(x_{N_i},0)-H(0,x_{N_j})+H(0,0)] \big\},
}
where, in our particular case,
\eq{
G(N_i,0) &= G(0,N_i) = G(0,0) = 0,
\\
H(N_i,0) &= H(0,0) = 0.
}
Then, substituting (\ref{Bsfortwo}) and (\ref{Csfortwo}), the form factor has two terms,
\eq{
f^{\ell\ell'} = \frac{g^2}{16\pi^2} s_{\nu_\ell}s_{\nu_{\ell'}}\,
\left[f^{(1)} + \sum_{k=1}^3 s_{\nu_k}^2\, f^{(2)}\right] ,
\label{f1f2}
}
where $f^{(1)}$ and $f^{(2)}$ do not depend on mixings. Only diagrams containing the flavor-changing vertex $H\chi_i\chi_j$ contribute to $f^{(2)}$, but we treat $G\chi\chi$ and $G\chi$ together since they cancel the ultraviolet divergences of each other, present in the part $f^{(1)}$. In the case of diagrams $G\chi\chi+G\chi$, the part $f^{(2)}$, subdominant at low neutrino masses in any case, cancels $f^{(1)}$ at some point and, for large neutrino masses, becomes the dominant contribution despite the $s_{\nu_k}^2$ suppression (see Fig.~\ref{fig2}). This is because it keeps growing like $m_{N_1}m_{N_2}$.

The case of one Dirac singlet (two equal-mass Majorana neutrinos) corresponds to: 
\eq{
f^{(1)} &= F(x_{N})+G(x_{N},x_{N})+H(0,x_N)-F(0),
\\
f^{(2)} &= H(x_{N},x_{N})-H(0,x_N).
}
The case of one active Dirac neutrino (sequential) can be recovered from:
\eq{
f^{\ell\ell'}_{\rm seq} &=  \frac{g^2}{16\pi^2} s_{\nu_\ell} s_{\nu_{\ell'}} 
\left[ F(x_{N})+G(x_{N},x_{N})-F(0) \right].
}


\end{document}